# Techno-Economic Analysis and Optimization of a Microgrid Considering Demand-Side Management

**Seyyed Danial Nazemi \***
*Department of Industrial and Systems Engineering*
*Rutgers University*
Piscataway, NJ, USA
danial.nazemi@rutgers.edu

**Khashayar Mahani**
*Quanta Technology, LLC*
Raleigh, NC, USA
mahani.khashayar@gmail.com

**Ali Ghofrani**
*Department of Industrial and Systems Engineering*
*Rutgers University*
Piscataway, NJ, USA
a.ghofrani@rutgers.edu

**Mahraz Amini**
*National Grid*
Waltham, MA, USA
mahraz.amini@nationalgrid.com

**Burcu E. Kose**
*Department of Industrial and Systems Engineering*
*Rutgers University*
Piscataway, NJ, USA
burcu.kose@rutgers.edu

**Mohsen A. Jafari**
*Department of Industrial and Systems Engineering*
*Rutgers University*
Piscataway, NJ, USA
jafari@soe.rutgers.edu

*Abstract*—The control and management of power demand and supply become very crucial due to the penetration of renewables in the electricity networks and energy demand increase in residential and commercial sectors. In this paper, a new approach is presented to bridge the gap between Demand-Side Management (DSM) and microgrid (MG) portfolio, sizing, and placement optimization. Although DSM helps customers to take advantage of recent developments in the utilization of Distributed Energy Resources (DERs) especially microgrids, a huge need for connecting DSM results to microgrid optimization is being felt. Consequently, we propose a novel model that integrates the DSM techniques and microgrid modules in a two-layer configuration. In the first layer, DSM is employed to minimize the electricity demand (e.g., heating and cooling loads) based on zone temperature set-point. Using the optimal load profile obtained from the first layer, all investment and operation costs of a microgrid are then optimized in the second layer. The presented model is based on the existing optimization platform developed by RU-LESS (Laboratory for Energy Smart Systems at Rutgers University) research group. As a demonstration, the developed model has been used to study the impact of smart HVAC control on microgrid compared to traditional HVAC control. The results show a noticeable reduction in the total annual energy consumption and the annual cost of microgrid

## I. INTRODUCTION

It is important to meet the future energy requirements without interfering with the environment and energy provision [1]. The reduction of fossil fuels at a rapid rate and the growth of its price make the emergence of green power markets necessary [2]. Currently, energy and regulation markets become very beneficial to own and operate microgrids. As well as renewable energy resources, applying efficient and novel Demand-Side Management (DSM) methods (e.g., demand response, energy efficiency techniques, and Distributed Energy Resources (DERs) instead of centralized fossil fuel power plants) help the society financially and environmentally. Demand response has potential benefits for building owners, both stand-alone and within a microgrid [3]. In general, DSM enables both energy consumers and suppliers to take advantage of recent developments in DERs, especially microgrids. DERs are controllable energy generation systems that are directly located near end-users or connected to the local distribution systems. Some benefits that could be obtained by integrating efficient DER systems would be power flow reduction, grid resiliency increase, and network investment reduction [4]. The microgrid encompasses a portion of an electric power distribution system that is located downstream of the distribution substation, and it includes a variety of DER units and different types of end users of electricity and/or heat [5]. Microgrids, based on the needs of the local network, can be operated in either grid-connected mode or islanded mode. They are seen to be increasingly important to achieve a reliable, flexible, and sustainable electricity network [6]. Every microgrid can include different electrical and thermal assets such as CHP systems, natural gas or diesel generators, renewable energy systems such as wind turbines and PV systems [7] with or without battery storage systems [8]–[10], boilers and chillers, etc. These assets help the thermal and electrical networks to generate required heat and electricity.

Related works to this study can be categorized under two topics, demand-side management and DER planning and microgrid optimization. The concept of demand-side management (DSM) for electric utilities has been a focus of studies in the past few decades [11], [12]. In a traditional power grid, proper metering may decrease demand response non-scheduled loads slightly [13]. However, in the smart grid, advanced metering infrastructures and novel DSM algorithms are more beneficial for suppliers and customers. Works related to DSM explore a broad range of techniques to optimize various resources including energy efficiency (EE), demand response (DR), and distributed generation (DG) to reduce peak demand. Zhou Wu et al. [14] propose a portfolio optimization method using different resource mixes such including "EE+DR", "EE+DG" and "EE+DR+DG". Another approach [15] uses binary particle swarm optimization to schedule demand-side resources. Newer approaches examine automated and smart techniques. [16] proposes a DSM methodology for connected buildings with the objective of energy-saving and load leveling. In [17], a game-theoretic approach is utilized to automate DMS, where the game consists of users trying to optimize their energy consumption schedule for a future smart grid. In [18], hybrid data-driven approaches were investigated to predict building indoor temperature response in order to be used for energy saving in demand response programs, pre- and post-occupy control strategies, building operation control optimization frameworks, and fault and failure detection. Likewise, [19] explores a game theory model for DSM with additional consideration to small energy storage capabilities. Another game-theoretic model for DSM considers the power supplier and the consumer to be contributors to a two-way game for resource optimization [20].

Strategies that involve optimization of DER differ significantly from conventional power systems and therefore require consideration of different factors [5]. In order to work effectively, optimal solutions for the design and schedule of DERs need to be considered, along with the volatility of the energy market. These topics make up most of the research done related to DER and microgrids. In [21], mixed-integer linear programming (MILP) is used to decide on the capacity, location, number, and type of a DER to optimize the total cost of investment, maintenance, and operation. Similarly, [22] addresses the problem of the optimal design and operational scheduling of a microgrid to be solved by a MILP formulation. Day-ahead scheduling comprises switching controllable appliances and determining the optimal input and output of the microgrid. The primary purpose of scheduling can vary from maximizing the social welfare of the microgrid [23] to minimize the total operation costs [24]. A case study done by [25] has shown that employing a combination of the mentioned methods is capable of reducing emissions and voltage and current variations significantly while showing that the electrical grid constraints have a powerful impact on DERs.

Besides considering different methods for DSM and DER optimization, design and utilization of efficient electronic machines and devices such as transformers in the network are required to be investigated in order to improve the efficiency of the distribution and transmission. Analysis of various transformer structures for low- and high-frequency applications is also necessary [26]. Moreover, one of the most important quantities in the design, performance, and maintenance of electronics machines and devices is the losses and its estimation procedure under different conditions [27].

In their review of many available DER optimization methods, Alarcon-Rodriguez et al. [4] have concluded that in the future, DER planning methods will have to consider controllable loads and DSM. This study aims to bridge the gap between DER and DSM by proposing an approach that considers both of them together.

## II. PROBLEM STATEMENT

This study deals with the impact of demand-side management techniques on microgrid portfolio, placement, and sizing optimization. Although some interesting works have been done on finding the optimal framework for DER systems, there is a gap in the coupling of the optimal demand load and DER systems optimization. In this regard, comprehensive methods have been employed to optimize residential and commercial building loads and enable them to use the available energy more efficiently without installing new transmission lines and generation assets. Then, the output of this model is fed to an optimization framework to design a microgrid integrated with power networks and district heating and cooling. Fig. 1 illustrates the proposed model.

This article provides optimal short-term planning for the operation of residential and commercial buildings and investigates its impact on long-term control and planning of a microgrid with several buildings. The demand-side management is done based on occupancy, accurate thermal models of building zones, and external signals, to fulfill environment comfort and lower electricity bills in cooling seasons [28].

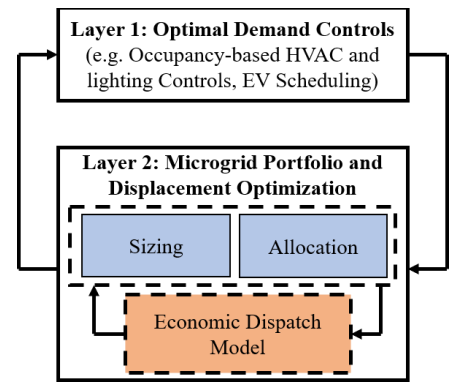

Fig. 1. The flowchart of the proposed model

The objective function of this problem is to minimize the ownership cost in an energy network using a mixed-integer linear programming model. The ownership cost is defined as the sum of energy costs and microgrid overall operation and investment costs. The proposed methodology could be applied to a broad range of urban systems, which consist of several nodes with different functionalities (residential, commercial, industrial, etc.). An overview of a general microgrid model is shown in Fig. 2.

Consider a network with 5 buses. In this network, there is a slack bus, which is connected to the main grid, and the other buses have controllable cooling systems. Each building has given thermal characteristics and human occupancy patterns. There are many indispensable factors for each individual zone that have different response patterns to the air conditioning system, such as size, internal gains, orientation, and occupancy patterns. This model considered the zonal thermal response, pre/post occupy HVAC operation, demand-side management programs, human comfort, and human productivity.

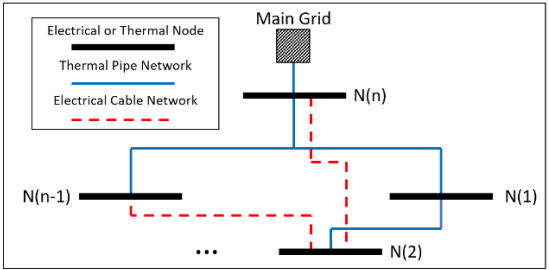

Fig. 2. The general topology of a microgrid with electrical and thermal networks

### III. MATHEMATICAL FORMULATION

The mathematical formulation for the integration of smart cooling control of buildings and optimal energy portfolio, sizing, and placement of microgrids is presented. The presented model is based on the existing optimization platform developed by RU-LESS (Rutgers University, Laboratory for Energy Smart Systems). This platform is being used for several projects in the state of New Jersey to deal with optimizing distributed energy systems. In this section, the formulation for each layer is separately presented. In the DSM part, the cooling load is optimized and will be the input of the next layer, which is microgrid portfolio optimization.

#### A. Smart HVAC Control

In order to find the optimal cooling load for residential and commercial buildings, the authors proposed a methodology that optimizes short-term planning for the cooling operation of buildings based on accurate thermal models of building zones to lower electricity bills in cooling seasons. The objective is to minimize the price for cooling load electricity, which comes from zone temperature setpoint and human performance penalty cost for comfort level deviation. The decision variable is temperature setpoint at time t for zone z ($T_z(t)$). There are also some constraints to be met namely the comfort zone temperature and the maximum temperature drop or rise at each time step [28]. Eq. 1 shows the objective function:

$$\min \sum_{z-1}^{N} \sum_{t-1}^{6} w_{z,1} ASE_z^t \times P^t + w_{z,2}(1 - HP_z^t) \times EP \quad (1)$$

where $w_{z,1}$ and $w_{z,2}$ are introduced to change the priority of cooling demand or human productivity in zone z, $P^t$ is the real-time energy price at time t in $/kWh, and EP is the constant profit made by each occupant at time t. In addition, $ASE_z^t$ and $HP_z^t$ are electricity consumption due to cooling load and relative human performance, respectively. $HP_z^t$ shows the relationship between the human comfort zone and task performance as follows [29]:

$$HP = 0.16475\,T - 0.00582\,T^2 + 0.00006\,T^3 - 0.46853 \quad (2)$$

where HP is relative human performance and T is dry-bulb temperature. Since a clear closed-form equation for finding the electricity consumption due to cooling load does not exist, a data-driven methodology was proposed and verified by the RU-LESS team to calculate it. Eq. 3 shows this multivariate linear regression model:

$$ASE_z^t = \beta_0 + \beta_1 \dot{E}_z^t + \beta_2 T_\infty^t + \beta_3 T_z^t + \beta_4 T_z^{t-1} + \beta_5 T_z^{t-2} + \beta_6 T_z^{t-3} + \beta_7 \tau^t + \beta_8 \sum_{i-1, i \neq z}^{N} T_i^t + \varepsilon^t \quad (3)$$

The above-mentioned equation illustrates that the cooling electricity consumption is a function of rate of change of energy in thermal zone (É), environment temperature ($T_\infty$), average zone temperature at time t, t-1, t-2, and t-3, time of the day which is assumed it is a categorical value ($\tau$), and the summation of average temperature for each zone at time t. Further details of this optimal model could be found in [28], which has been published before by the RU-LESS team.

#### B. Microgrid Portfolio and Displacement Optimization

In this section, a mathematical formulation is presented to address techno-economic analysis and optimization of a microgrid with electrical and thermal networks. The objective function of this model is set to minimize the total microgrid investment and operation costs. Eq. 4 shows the objective function of this model:

$$\min\{C_{invd} + C_{invc} + \sum_{y=1}^{M}(C_{pur}^y + C_{dem}^y + C_{gen}^y - C_{exp}^y)\} \quad (4)$$

The objective function includes all operation and investments costs of a microgrid, namely: the investment cost of discrete ($C_{invd}$) and continuous ($C_{invc}$) technologies, and the summation of cost of electricity purchase ($C_{pur}$), the demand charges ($C_{dem}$), the generation costs ($C_{gen}$), and the revenues from electricity export ($C_{exp}$) from the first year until year M when is the planning horizon.

In this study, the bus voltages in meshed/radial distribution networks are approximated by a linearization algorithm proposed by Bolognani and Zampieri [30]. In the proposed model, the net injected power at each node is the main constraint, which is the summation of local generation at that node, utility import or export, battery charging or discharging, load and load curtailment, and electric chiller consumption.

In this model, other constraints are also considered for cable currents, bus voltages, energy generation, and energy storage systems by using the MILP model proposed by Mashayekh et al. [31]. In their model, they linearized the equations for cable current constraints. Also, they enhanced the approach originally proposed by Franco et al. [32] to linearize the bus voltage constraints.

Moreover, the heating and cooling balances are taken into account at each node using the linear approximation of thermal losses proposed by Söderman and Pettersson [33]. To calculate the heat balance at each node, the summation of heating loads and resources, recovered heat from CHP units, charging or discharging of heat storage systems, the needed heat for absorption chilling and the heat transfer between nodes are considered. For finding cooling energy at each node, cooling loads, cooling energy from absorption chilling, and cooling energy from cold storage systems are taken into account.

## IV. CASE STUDY

### A. Smart Cooling Control Results and Microgrid Setup

The proposed model was applied to a 5-node microgrid. This 12 kV microgrid contains 4 buildings (2 medium-sized office buildings and 2 midrise apartments) shown in Fig. 3. In this study, we consider two scenarios. In the first scenario, building load profiles were generated using EnergyPlus reference buildings and cooling systems have ordinary setback control systems. The second scenario consists of building load profiles after applying the smart cooling control. The results of both scenarios are fed to the microgrid portfolio, sizing, and placement optimization model to illustrate the impact of demand-side management techniques on the microgrid for the same design problem.

Table 1 shows annual electrical (all electrical loads except cooling) and cooling load for each node. For each one, two case scenarios are considered. The annual loads for the first scenario come from Energy Plus for each building; however, the annual loads for the second scenario are the results of applying smart cooling controls to the buildings.

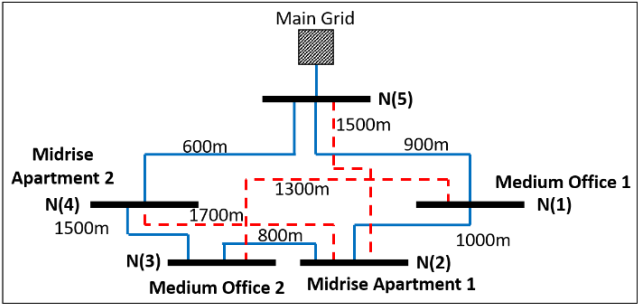

Fig. 3. The 5-Node Microgrid Network

TABLE I.  ANNUAL ELECTRICAL AND COOLING DEMAND

| Node | Case | Annual Electrical Load | | Annual Cooling Load | |
|---|---|---|---|---|---|
| | | Usage (MWh) | Peak Demand (kW) | Usage (MWh_th) | Peak Demand (kW_th) |
| 1 | 1 | 1819 | 446 | 12958 | 2442 |
| | 2 | 1819 | 446 | 11223 | 2088 |
| 2 | 1 | 1399 | 297 | 1497 | 815 |
| | 2 | 1399 | 297 | 1293 | 688 |
| 3 | 1 | 1450 | 452 | 2055 | 1196 |
| | 2 | 1450 | 452 | 1752 | 1006 |
| 4 | 1 | 644 | 211 | 233 | 622 |
| | 2 | 644 | 211 | 202 | 533 |

Furthermore, in this study, the microgrid owner is allowed to invest in Energy Storage System (ESS), PV system, and also CHP system with either Micro Turbine or Fuel Cell as its prime mover. The characteristics of each above-mentioned technology are presented in Table 2.

TABLE II.  ANNUAL ELECTRICAL AND COOLING DEMAND

| Discrete Technology | | | | |
|---|---|---|---|---|
| Technology | Capacity (kW) | Capital Cost ($/kW) | Lifetime (years) | Efficiency |
| CHP with MT | 2000 | 3500 | 25 | 41% |
| CHP with FC | 1000 | 4000 | 25 | 37% |
| Continuous Technology | | | | |
| Technology | Fixed Cost ($/kW) | Variable Cost ($/kW) | Lifetime (years) | |
| ESS | 600 | 500 | 6 | |
| PV | 3000 | 2000 | 25 | |

In the electrical network, cables have the ampacity of 0.4 pu and the impedance of $6 \times 10^{-6}$ pu/m. Also, PJM rate structures are used to calculate the real-time price of electricity.

### B. Microgrid Portfolio and Sizing Results

Table 3 illustrates the annual investment and operation costs for each case scenario. The results show that by applying smart cooling control, the microgrid investment cost has decreased by 13.79%. The annual microgrid operation cost has also dropped to 148,346 $/year, which shows a decrease of 9.15%. In addition, the total annual cost has reduced by 10.67%, which is a noticeable decrease.

Table 4 shows the optimal sizing and placement of different technologies at each node for both scenarios. By comparing these two case scenarios, it can be understood that the aggregate technology capacities will decrease significantly. Furthermore, in both cases, CHP units have the same capacity. On the other hand, PV and ESS capacities are different for each case, and applying small HVAC control reduces the required capacity of PV and ESS.

TABLE III.  MICROGRID PORTFOLIO AND SIZING RESULTS

| Case Scenario | Investment Cost ($) | Annual Operation Cost ($) | Total Annual Cost ($) |
|---|---|---|---|
| **Scenario I** (Setback Control for HVAC) | 79,420 | 163,291 | 242,711 |
| **Scenario II** (Smart Control for HVAC) | 68,461 | 148,346 | 216,807 |

TABLE IV.  OPTIMAL TECHNOLOGY SIZING AND DISPLACEMENT (ALL NUMBERS ARE IN KW

| Node | Case | PV | CHP (MT) | CHP (FC) | ESS |
|---|---|---|---|---|---|
| 1 | 1 | - | 2000 | - | - |
| | 2 | - | 2000 | - | - |
| 2 | 1 | 680 | - | - | 200 |
| | 2 | 550 | - | - | 140 |
| 3 | 1 | - | - | 1000 | - |
| | 2 | - | - | 1000 | - |
| 4 | 1 | 350 | - | - | 120 |
| | 2 | 300 | - | - | 100 |
| Aggregate | 1 | 1030 | 2000 | 1000 | 320 |
| | 2 | 850 | 2000 | 1000 | 240 |

Fig. 4 and 5 show the optimal electricity dispatch for both case scenarios at Node 1 and Node 2 for a typical weekday in mid-July. Figure 4 illustrates that the CHP unit provides a considerable share of the required electricity and cooling demands to medium office 1. By comparing these two figures can also be understood that applying smart cooling control results in peak reduction and decrease in the imported electricity from the grid.

Figure 5 illustrates the hourly dispatch at midrise apartment 1. It can be seen that the PV system provides a reasonable share of electricity and cooling demands during the day. In addition, the energy storage system is charged during off-peak hours and discharged during peak hours to help the network be less dependent on the grid. These two plots in Figure 5 also present that Node 2 imports the electricity from other nodes, especially those nodes with CHP units, to reduce its need to get electricity from the main grid.

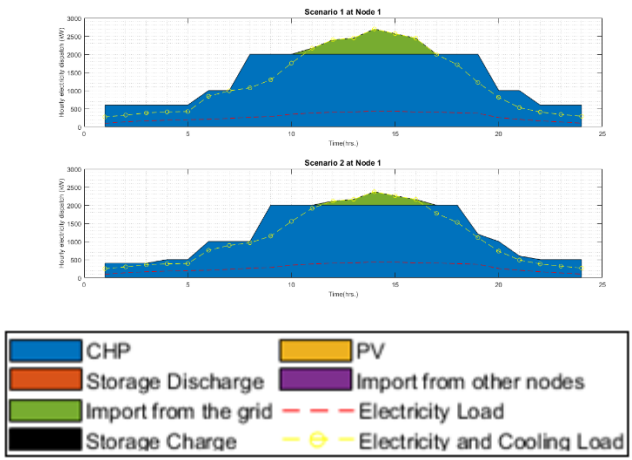

Fig. 4. Optimal electricity dispatch for both scenarios at Node 1

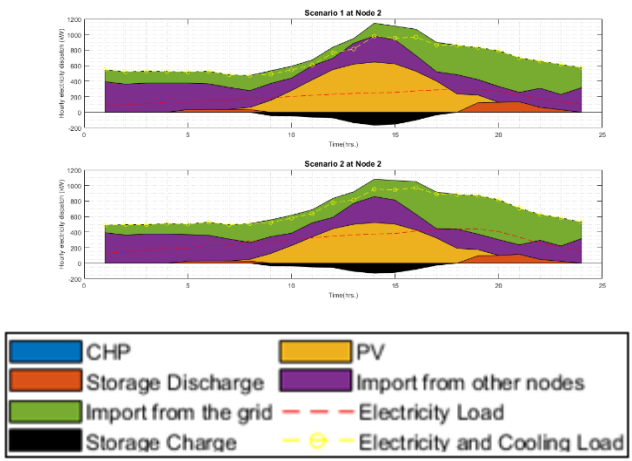

Fig. 5. Optimal electricity dispatch for both scenarios at Node 2

## V. CONCLUSION

This article presented the impact of a demand-side management technique on microgrid portfolio, sizing, and placement. The proposed model consisted of two main layers, namely: smart HVAC control and microgrid portfolio optimization. The main objective functions of this work were minimizing the price for cooling load electricity, which comes from zone temperature setpoint and human performance penalty cost for comfort level deviation, and optimizing all operation and investment costs of a microgrid using the results of the demand-side management model. To show the impact of smart HVAC control on the microgrid, an illustrative example was investigated. This example showed that the total annual cost of microgrid reduced by 10.67% when a smart cooling control was applied to each node. The results of the proposed model and its implementation on the example illustrated that the performance of a microgrid could even be improved further by building new bridges between demand-side techniques and microgrid management.